\documentclass[format=manuscript, review=false, screen=true]{acmart}

\usepackage{booktabs} 

\usepackage[ruled]{algorithm2e} 

\SetAlFnt{\small}
\SetAlCapFnt{\small}
\SetAlCapNameFnt{\small}
\SetAlCapHSkip{0pt}
\IncMargin{-\parindent}

\usepackage{listings} 
\definecolor{verylightgrey}{gray}{0.9}
\usepackage{amssymb}
\usepackage[flushleft]{threeparttable}

\usepackage{graphicx}
\usepackage{subcaption}
\graphicspath{{images/}}

\usepackage{longtable}
\usepackage{multirow, makecell}
\usepackage{rotating}

\acmJournal{TECS}
\acmVolume{18}
\acmNumber{3}
\acmArticle{21}
\acmYear{2019}
\acmMonth{4}
\copyrightyear{2019}

\setcopyright{acmlicensed}

\acmDOI{10.1145/3319618}

\begin{document}
\title[A Survey of Asynchronous Programming Using Coroutines in IoT and Embedded Systems]{A Survey of Asynchronous Programming Using Coroutines in the Internet of Things and Embedded Systems}

\author{Bruce Belson}
\orcid{0000-0003-2904-1650}
\affiliation{%
  \institution{James Cook University}
  \department{College of Science \& Engineering}
  \city{Cairns}
  \state{Queensland}
  \postcode{4870}
  \country{Australia}}
\email{bruce.belson@my.jcu.edu.au}

\author{Jason Holdsworth}
\affiliation{%
	\institution{James Cook University}
	\department{College of Science \& Engineering}
}
\email{jason.holdsworth@jcu.edu.au}

\author{Wei Xiang}
\affiliation{%
	\institution{James Cook University}
	\department{College of Science \& Engineering}
}
\email{wei.xiang@jcu.edu.au}

\author{Bronson Philippa}
\affiliation{%
	\institution{James Cook University}
	\department{College of Science \& Engineering}
}
\email{bronson.philippa@jcu.edu.au}

\thanks{%
Authors' addresses: Bruce Belson, James Cook University, College of Science \& Engineering, Cairns, Queensland, 4870, Australia, bruce.belson@my.jcu.edu.au; Jason Holdsworth, James Cook University, College of Science \& Engineering, jason.holdsworth@jcu.edu.au; Wei Xiang (corresponding author), James Cook University, College of Science \& Engineering, wei.xiang@jcu.edu.au; Bronson Philippa (corresponding author), James Cook University, College of Science \& Engineering, bronson.philippa@jcu.edu.au.\\}


\begin{abstract}
Many Internet of Things and embedded projects are event-driven, and therefore require asynchronous and concurrent programming. Current proposals for C++20 suggest that coroutines will have native language support. It is timely to survey the current use of coroutines in embedded systems development.  This paper investigates existing research which uses or describes coroutines on resource-constrained platforms. The existing research is analysed with regard to: software platform, hardware platform and capacity; use cases and intended benefits; and the application programming interface design used for coroutines. A systematic mapping study was performed, to select studies published between 2007 and 2018 which contained original research into the application of coroutines on resource-constrained platforms. An initial set of 566 candidate papers, collated from on-line databases, were reduced to only 35 after filters were applied, revealing the following taxonomy. The C \& C++ programming languages were used by 22 studies out of 35. As regards hardware, 16 studies used 8- or 16-bit processors while 13 used 32-bit processors.  The four most common use cases were concurrency (17 papers), network communication (15), sensor readings (9) and data flow (7). The leading intended benefits were code style and simplicity (12 papers), scheduling (9) and efficiency (8). A wide variety of techniques have been used to implement coroutines, including native macros, additional tool chain steps, new language features and non-portable assembly language. We conclude that there is widespread demand for coroutines on resource-constrained devices. Our findings suggest that there is significant demand for a formalised, stable, well-supported implementation of coroutines in C++, designed with consideration of the special needs of resource-constrained devices, and further that such an implementation would bring benefits specific to such devices.
\end{abstract}


%
%


\maketitle
\thispagestyle{empty}


\section{Introduction}

The Internet of Things (IoT) \cite{Al-Fuqaha2015, Atzori2010, Gubbi2013} continues to grow both in the scale and variety of attached devices and in the number of developed applications \cite{Manyika2015, VanderMeulen2017}. This growth draws attention to the software engineering of the resource-constrained embedded systems that are a frequent component of heterogeneous IoT applications. Such attention is all the more urgently required because of new challenges with regard to security \cite{Sicari2015}, reliability \cite{Gubbi2013} and privacy \cite{Weber2015}.

Many IoT and embedded systems have an event-driven architecture; their software is consequently implemented in an asynchronous programming style, whereby multiple tasks wait on external events. Asynchronous code is challenging to write because application logic becomes split between the function initiating the request and the event handler that is invoked when the response is ready \cite{Gay2003, Levis2002, Meijer2010}. This "split-phase" architecture becomes increasingly complex when the developer introduces more event sources (such as timeouts) with their own event handlers. There may be interaction between various split-phase events, which can add degrees of freedom to the various state models: consequently there is an increasing likelihood that the source code addressing a single event is split between separate locations, forcing the reader to jump between them. Application logic is obscured by the split-phase fragmentation, leading to a gap between the design of the system and its source code representation, making the code harder to understand and more difficult to maintain \cite{Brodu2015, Edwards2009, Madsen2017, Kambona2013}.

A solution to the split-phase problem for desktop software has been language support for coroutines \cite{Conway1963, Knuth1968, Marlin1979} and promises \cite{Brodu2015, Liskov1988, Madsen2017}. For example, in C\#, JavaScript, and Python, developers can use an "await" keyword to wait on an external event. This means that asynchronous code can be written just as clearly as the equivalent code in a synchronous style that uses blocking code. However, resource-constrained embedded systems are overwhelmingly programmed in C or C++ \cite{AspenCore2017, Skerrett2017}, which lack support for the "await" pattern.

The C++ standardisation committee is currently debating the inclusion of coroutines, and at least two competing designs have been proposed \cite{ISO2017, Romer2018}. The addition of coroutines to C++ would create an opportunity to simplify embedded systems code. Existing research on coroutines in C++ may not have considered the needs of embedded systems and other extremely resource-constrained devices, because the initial implementations used compilers that do not target such platforms \cite{Mittelette2015}. Here, we specifically focus on small embedded systems that have insufficient memory to run Linux or another general purpose OS. If the C++ language adds the async/await and coroutine patterns, we believe it is important that the needs of resource-constrained platforms are also considered.

This article contains a systematic mapping of the use of coroutines in embedded systems, conducted by searching academic databases and manually inspecting every matching paper. It thus provides a complete perspective on academic research addressing the use of coroutines in embedded systems to inform the C++ standardization process by identifying how and why coroutines are used. The article uses the mapping to build a taxonomy of existing research with regard to platform, use cases and implementation.

The design of the study, details of the methodology used for each stage and the results of each stage are available in spreadsheet format.
The remainder of this paper is organised as follows. Section \ref{section:background} contains the background, beginning with an introduction to the development environment for C/C++ programs on resource-constrained devices, to some of the problems commonly encountered by developers and to the types of solution currently applied to these problems. It continues with a discussion of the use of coroutines in C and C++. Section \ref{section:sms} details the methodology of the mapping process used in this study, some of the logic underpinning the methodological choices, and a review of related work. Section \ref{section:results} explores the results and presents insights. Section \ref{section:analysis} contains a discussion of results and an analysis of research gaps. Section \ref{section:conclusion} discusses future research possibilities and concludes the paper.

\section{Background}
\label{section:background}

\subsection{Async/Await pattern}

Much of the program flow in IoT and embedded device programs is asynchronous, for example, requiring the software to wait on responses from a remote device or sensor. A naïve approach to implementing this flow results in complex arrangements, such as a finite state machine (FSM) and multiple fragments of code. This produces source code that is complex, fragile and inflexible.

Alternatively, there are two common patterns for a simpler and more robust design. The first, continuation-passing style (CPS), which is seen commonly in JavaScript, can lead to the "pyramid of doom" or "callback hell" phenomenon \cite{Brodu2015, Edwards2009, Madsen2017, Kambona2013} when multiple sequential operations are composed.

A more elegant approach is the async/await pattern \cite{Bierman2012, Haller2016, Okur2014, Syme2011}, which is a popular device for transforming continuation-passing style code into direct programming style, with all the asynchronous steps of a sequence written in a single ordered sequence within a single block of code. The pattern has been used successfully in several languages, notably C\# \cite{Bierman2012, Okur2014} and JavaScript, as part of the ECMAScript 2018 proposal \cite{ECMA2017}; in C++, proposals are currently being considered for inclusion in the C++ 2020 standard, using new keywords 'co\_await', 'co\_yield' and 'co\_return' or alternative syntax \cite{ISO2017, Romer2018}. The async/await pattern allows the programmer to write a single continuous set of statements in a direct programming style, which will be performed in the correct order, even when they are run asynchronously as a set of separate events. Furthermore, the pattern avoids the explicit use of global variables.

\subsection{Coroutines}

Coroutines extend the concept of a function by adding suspend and resume operations \cite{Conway1963, Knuth1968, Marlin1979}. Coroutines can be used for a variety of purposes including (i) event handlers \cite{Dunkels2006}; (ii) data-flow \cite{Kugler2013}; (iii) cooperative multitasking \cite{Susilo2009} as well as (iv) the async/await pattern \cite{ISO2017}.

During suspension, the implementation stores the execution point of the coroutine, and usually (but not always) the state of local variables. For example, Protothreads \cite{Dunkels2006} is a coroutine implementation for embedded systems where local variables are not preserved: instead all variables within the coroutine must be statically allocated. This strategy reduces the overhead of context switching and provides predictable memory usage but produces coroutines that are non-reentrant. Furthermore, code defects are more likely when the programmer is responsible for explicitly managing coroutine state. This study will examine both types of coroutine in the context of embedded systems.

Coroutine implementations may be further categorised into stackful or stackless types. A stackful coroutine has its own stack which is not shared with the caller, and hence local variables can be stored there during suspension. Conversely, a stackless coroutine pops its state off the stack during suspension (like a normal function return). For stackless coroutines, other mechanisms must be introduced in order to preserve state, such as storing local variables in global storage. 

Furthermore, a stackless coroutine often can only be suspended from within the coroutine itself and not from a subroutine (i.e. a function called from the coroutine). For example, C++ proposal N4680 is a stackless model that requires all yield or return statements to be contained within the body of the coroutine.

Neither model is considered universally appropriate for the various C++ use cases \cite{Goodspeed2014, Riegel2015}. Alternative techniques, such as stack slicing, have been used to preserve state in a stackless implementation and provide single threaded cooperative multitasking \cite{Tismer2000, Tismer2018}.

\subsection{Previous coroutine implementations for constrained platforms}
\label{previous-implementations}

Early implementations used macros in C to add coroutine-like features. For example, Duff's device takes advantage of the fall-through behaviour of C's case statement in the absence of a break statement \cite{Duff88}. It is unusual in that a block such as do … while, can be interleaved within the case statements of a switch statement. \citeN{Tatham2000} described a coroutine solution in C, which makes use of Duff's device to efficiently implement coroutines through macros, without the need to explicitly code a state machine. However, Tatham noted that "this trick violates every coding standard in the book" and Duff called the method a "revolting way to use switches to implement interrupt driven state machines". This technique was extended by Dunkels et al. for Protothreads \cite{Dunkels2006}, which provided conditional blocking operations on memory-constrained systems, without the need for multiple stacks, and formed the core of the widely used real-time operating system Contiki \cite{Dunkels2004}.

Protothreads (and any other solution based on Duff's device) can be considered to suffer from two serious defects. First, their use adds a serious constraint to C programs: switch statements cannot be used safely in programs that use Protothreads; they may cause errors that are not detected by the compiler but cause unpredictable behaviour at run-time. Second, they do not manage local variable state on behalf of the programmer: any variable within the coroutine whose state should be maintained between calls must be declared as static (global) \cite{Dunkels2005b}. This has consequences for reentrancy, and for code quality. On the other hand, they are an extremely cheap solution in terms of coding effort, memory use and speed, and they are portable, because they use pure C. The original library is written in C; it has been ported to C++ \cite{Paisley2006}.

Listing \ref{protothreads} contains a fragment of code that used Protothreads to implement part of an asynchronous producer/consumer pattern. Listing \ref{coawait} shows a similar code fragment, this time using C++ language features, including the co\_await keyword of the current C++ standardisation proposal N4680. We observe several differences between the two. Listing \ref{coawait} contains fewer lines of code than Listing \ref{protothreads}; Listing \ref{coawait} does not contain macros; Listing \ref{protothreads} requires that local variables be declared as 'static', but Listing \ref{coawait} does not. While both code fragments present conceptual changes from the synchronous equivalent, we believe that the change in Listing \ref{coawait} is more transparent and more clearly presented.

\begin{lstlisting}[language=C, caption=Fragment of Protothreads code for asynchronous producer/consumer threads, label=protothreads, float=t]
static struct pt_sem full, empty;

static
PT_THREAD(consumer(struct pt *pt))
{
  static int consumed;
  PT_BEGIN(pt);
  for (consumed = 0; consumed < NUM_ITEMS; 
      ++consumed) {
    PT_SEM_WAIT(pt, &empty);
    consume_item(get_from_buffer());
    PT_SEM_SIGNAL(pt, &full);
  }
  PT_END(pt);
}
\end{lstlisting}

\begin{lstlisting}[language=C++, caption=C++ code fragment using co\_await for asynchronous producer/consumer threads, label=coawait, float=t]
task<> consumer(semaphore& sem) {
  auto producer = async_producer(sem, NUM_ITEMS);
  for co_await(auto consumed : producer) {
    consume_item(get_from_buffer());
  }
}
\end{lstlisting}

In 1992, Gupta et al. examined a coroutine-based concurrency model for resource-constrained platforms as part of a comparison between alternative models \cite{Gupta1992}. In 2000, Engelschall summarised the various techniques based on setjmp \& longjmp \cite{Engelschall2000}. FreeRTOS \cite{Barry2018} is an open source real-time operating system developed "around 2003" that contains a coroutine scheduler: local variable state is not maintained. In 2006, Rossetto and Rodriguez described a new concurrency model \cite{Rossetto2006} implemented as an extension to TinyOS \cite{Levis2005}, using coroutines as the basis of the integration; the implementation is stackful and local variables' states are maintained. 


\citeN{Schimpf2012b} provides a modified version of Protothreads which supports a priority-based scheduler. \citeN{Cohen2007b} provide a coroutine-based scheduler for TinyOS \cite{Levis2005} which is used to implement "RPC-like interfaces"; these support a direct programming style for communications code written in nesC \cite{Gay2003}. \citeN{Riedel2010} generate C code for multiple platforms, including a version that uses coroutines to provide concurrency. \citeN{Susilo2009} use a coroutine-based scheduler to achieve "[r]eal time multitasking […] without interrupts". Finally, \citeN{Andersen2017b} reject the use of C++ futures because the implementation model needs to handle a stream of events, rather than a single event, and is therefore non-deterministic in its use of memory, which is undesirable on a constrained platform: "One is forced therefore to trade off the reliability of promises […] in order for them to work in the embedded space." Instead, the authors use callbacks for C++ event handling.

The scripting language Lua possesses a coroutine implementation \cite{Moura2004a} and has been successfully used on microcontrollers \cite{Hempel2008}. MicroPython \cite{George2014a} is a Python 3 version which supports microcontrollers \cite{George2014b} and includes support for generators and coroutines \cite{VanRossum2005, Python2017}.

\subsection{Programming Languages: C and C++}

The majority (78\%) of embedded systems are programmed in C or C++ according to the 2017 Embedded Markets Study \cite{AspenCore2017}. The C language is the most popular, but its usage is slowly declining over time in favour of C++ and other languages. Between 2015 and 2017 the proportion of embedded projects using C++ rose from 19\% to 22\%, while C use fell from 66\% to 56\%. Coroutines are proposed for the C++ language, not C, so embedded programmers would need to use C++ to access these features. We believe that C++ usage will continue to increase, and therefore the design of C++ coroutines should consider the constraints of embedded software.  

The switch from C to C++ need not be dramatic. C++ is close to being a superset of C \cite{Stroustrup1986}. With the right compiler support, it is possible to migrate an embedded code-base from C to C++ merely by changing a compiler flag. There are potential problems with the migration from C to C++, including the possibilities that the code produced may be larger, slower and less likely to contain blocks that are appropriate for placement in ROM than the C code \cite{Goldthwaite2006, Herity2015}. A more subtle problem is that the code may be less amenable to worst-case execution time (WCET) analyses. Goldthwaite \cite{Goldthwaite2006} examined these problems, and identified three areas where difficulties might exist despite defensive programming, all of them with regard to timing analysis: (i) dynamic casts, (ii) dynamic memory allocation and (iii) exceptions.

The features that might persuade a development team to make the move to C++ have always included well-known front-end features such as namespaces, encapsulation, and inline functions, all of which offer benefits regarding code clarity but have no implementation cost in terms of code size or speed. Replacing split-phase functions with a direct programming style by using new, widely supported, language standards would appear to be a strong enticement for developers to migrate. It remains to be seen whether the feature can be provided for embedded systems without including two of the three behaviours which Goldthwaite \cite{Goldthwaite2006} identified as being problematical: dynamic memory allocation and exceptions.

\section{Systematic mapping study}
\label{section:sms}

\begin{figure}[h]
	\includegraphics[width=0.95\textwidth]{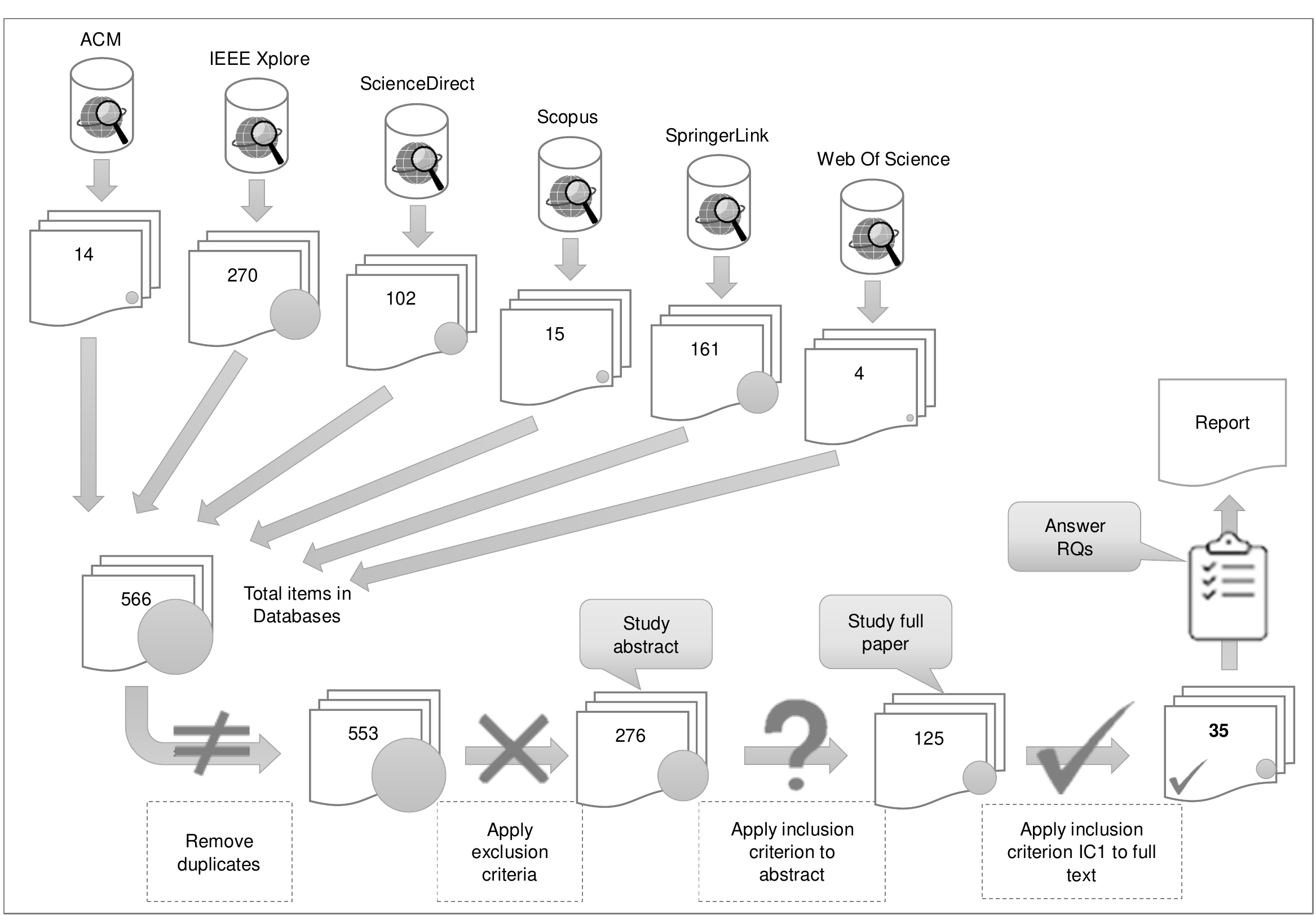}
	\caption{Summary of the search and selection process}
	\label{fig:sms}
\end{figure}

\subsection{Overview}

This systematic mapping study is informed by the guidelines of Kitchenham \cite{Kitchenham2004}, Kitchenham and Charters \cite{Kitchenham2007} and Petersen et al. \cite{Petersen2008}. The process is illustrated in Figure \ref{fig:sms}. The process searched six online databases, selected for relevance \cite{Brereton2007} and availability, for papers containing a term from each of the lists in Table \ref{table-databases}. We ensured completeness by iterative testing using snowballing \cite{Kitchenham2011a, Petersen2015} and by careful handling of database-specific behaviours regarding plurals, spellings and abbreviations.

\begin{table}[h]%
	\caption{Search strings used for online databases}
	\label{table-databases}
	\begin{tabular}{ p{4.5cm} c p{7cm} }
		\hline
		Part 1: Pattern & & Part 2: Platform\\
		\hline
		coroutine OR "lightweight thread" & AND & IoT OR "Internet of Things" OR "Cyber Physical Systems" OR RTOS OR "Real-time Operating Systems" OR "Embedded Systems" OR WSN OR "Wireless sensor networks" OR WSAN\\
		\hline
	\end{tabular}
\end{table}

\subsection{Search procedure}

The main inclusion criterion was that the paper should contain original research into the application of coroutines on resource-constrained platforms (IC1). This criterion excluded a large body of papers which applied coroutines only within the simulation of resource-constrained platforms, not on the platform itself.

The exclusion criteria were informed by previous studies \cite{Kitchenham2007, Petersen2008}. Papers were excluded if they lacked a scholarly identifier such as DOI or ISBN (EC1) or an abstract (EC2), were published before 2007 (EC3), were not written in English (EC4), were not available to the reviewers (EC5), were earlier versions of another paper (EC6), were not primary studies (EC7) or were not in any of the selected publication classes (journal articles, conference papers or book chapters) (EC8).

Two searches were conducted in October 2017 and in September 2018 across all databases. These searches resulted in 187 journal articles, 224 conference papers and 155 book chapters. This informed our decision to include all three publication classes within the search domain. The decision was made to include only studies published since 2007; this criterion excluded approximately 43\% of the original search results.

Details of the search strings, inclusion and exclusion criteria, procedures and download scripts can be found in the supplementary materials.

\subsection{Other systematic reviews and mapping studies}

An initial tertiary study was executed, being a review of existing reviews and mapping studies in the area of interest, as suggested by Kitchenham and Charters' guidelines \cite{Kitchenham2007}. The work concluded that, at the time of writing, this study appears to be the first to systematically map the use of coroutines in resource-constrained systems, whether embedded systems or IoT component systems.

\subsection{Research questions}

A major motivation for the study was to prepare the ground for an acceptable implementation of the await/async and generator patterns on resource-constrained platforms, using coroutines. The research questions therefore address what is known about hardware and software platforms, developer preferences, use cases, intended benefits, and application programming interfaces (APIs).

RQ1 investigated the software platform, including the programming language, the operating system and the implementation used for the relevant language feature. RQ2 looked at the hardware platform, including memory size and processor family. RQ3 and RQ4 assessed the use cases and intended benefits respectively of the coroutine usage. RQ5 assessed the programming interface.

The research questions are listed in full in Table \ref{RQs}. By examining the hardware and software characteristics of previous implementations we aimed to identify the salient characteristics of the environment within which a coroutine implementation must function. By investigating use cases and desired outcomes, we would identify some of the necessary characteristics of a successful implementation. Finally, by examining the programming interface we hoped to observe how researchers addressed some of the design trade-offs of the implementation.

\begin{table}[h]
	\caption{Research questions}
	\label{RQs}
	
	\begin{tabular}{l l l}
		\hline
		Code & & Research question \\
		\hline
		RQ1	& & What was the software platform?\\
		& RQ1a & What was the programming language used?\\
		& RQ1b & What method was used to implement coroutines?\\
		& RQ1c & What was the operating system used (if any)?\\
		RQ2	& & What was the hardware platform?\\
		& RQ2a & What was the class of hardware platform?\\
		& RQ2b & How much read-only or flash memory (ROM) was available?\\
		& RQ2c & How much random-access memory (RAM) was available?\\
		& RQ2d & What was the processor family?\\
		& RQ2e & Was the processor 8-bit, 16-bit or 32-bit?\\
		& RQ2f & What was the processor's instruction set?\\
		RQ3	 &  & What were the use cases?\\
		RQ4	 &  & What were the intended benefits of using coroutines in this context?\\
		RQ5	 &  & What is the API of the coroutine?\\
		& RQ5a & Does the paper discuss an implementation of coroutines?\\
		& RQ5b & Is the control flow managed on behalf of the developer?\\
		& RQ5c & Is the state of local variables automatically managed?\\
		& RQ5d & Is the coroutine implementation stackless or stackful?\\
		& RQ5e & How is the coroutine state allocated?\\
		\hline
	\end{tabular}
\end{table}

\subsection{Threats to validity}

Data extraction followed the principles laid down in Petersen et al. \cite{Petersen2015} for repeatability.

The validity of the results of this study are exposed to multiple sources of threat, particularly with regard to (i) study selection, (ii) data extraction and (iii) classification.

During study selection, the search process was recorded in detail and the search strings were tested for repeatability and for consistency across databases. Snowballing describes the process of expanding the search results by recursively selecting papers cited by a selected paper or which cite a selected paper \cite{Kitchenham2011a, Petersen2015}. While the study did not utilise snowballing during the final search process, it did use it during the earlier stages of establishing search strings, and some searches were consequently amended. During the application of selection criteria, the reviewers conferred whenever differences arose, and periodically discussed and reviewed the processes being used, using both contentious cases and randomly selected test papers to compare individual processes.

The guidelines of Petersen et al. \cite{Petersen2008, Petersen2015} were followed with regard to the data extraction process: a data collection form was constructed in Excel, and was used consistently to record the process, in order to improve repeatability and accuracy, and to reduce subjectivity.

To improve the consistency of classification a subset of papers was inspected by both reviewers, and the classifications were compared and discussed. This comparison was iterated until the rationale for classifications was fully established and any contentious cases had been resolved.

\subsection{Data set}

The initial search found 566 results; removal of duplicates left 553 documents. More than half of these failed the exclusion criteria, leaving 276 whose abstracts were studied.

Approximately 55\% of the surviving studies immediately failed the inclusion criteria, leaving 125 to be studied in full. After applying the inclusion criteria on the basis of the entire text, about 72\% failed, and 35 studies were retained \cite{Alvira2013, Andalam2014, Andersen2016, Andersen2017b, Bergel2011, Boers2010, Clark2009, Cohen2007b, Durmaz2017, Elsts2017, Evers2007, Fritzsche2010, Glistvain2010, Inam2011, Jaaskelainen2008, Jahier2016, Kalebe2017, Karpin2007, Khezri2008, Kugler2013, Kumar2007, Liu2011, Lohmann2012, Motika2015, Niebert2014, Noman2017, Oldewurtel2009, Park2015, Riedel2010, Schimpf2012b, St-Amour2010, Strube2010, Susilo2009, VonHanxleden2009, Yu2008}. Of these, 21 studies included a discussion of the implementation of coroutines. The lower half of Figure \ref{fig:sms} illustrates the process.

The selected papers addressed the issue of coroutines despite the lack of mainstream language support. These researchers identified a need that was not addressed by common languages and showed the potential benefits of these features. Now that native asynchronous programming support is being added to the C++ language, it is likely that demand from embedded software developers will only increase.

\section{Results}
\label{section:results}

\subsection{Overview}

The research identified 35 papers of relevance, of which 21 described coroutine implementations, developed in 7 different programming languages. Detailed lists of the results may be found in the supplementary materials.

\subsection{Programming language}

\begin{figure}[h]
	\centering
	\begin{subfigure}[h]{0.48\textwidth}
		\centering
		\includegraphics[width=\textwidth]{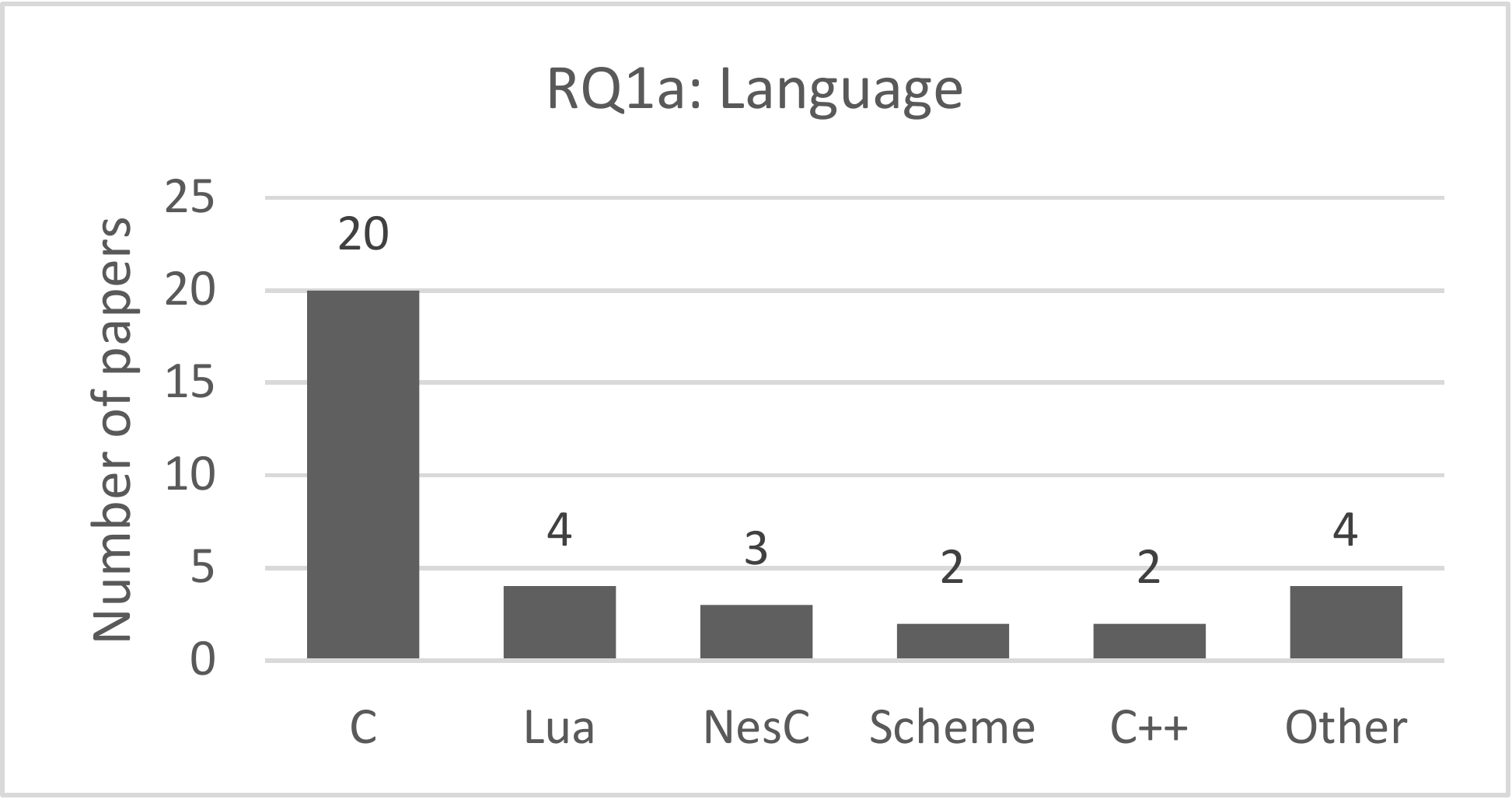}
		\caption{RQ1a - Programming languages used}
		\label{fig:rq1a}
	\end{subfigure}
	\begin{subfigure}[h]{0.46\textwidth}
		\centering
		\includegraphics[width=\textwidth]{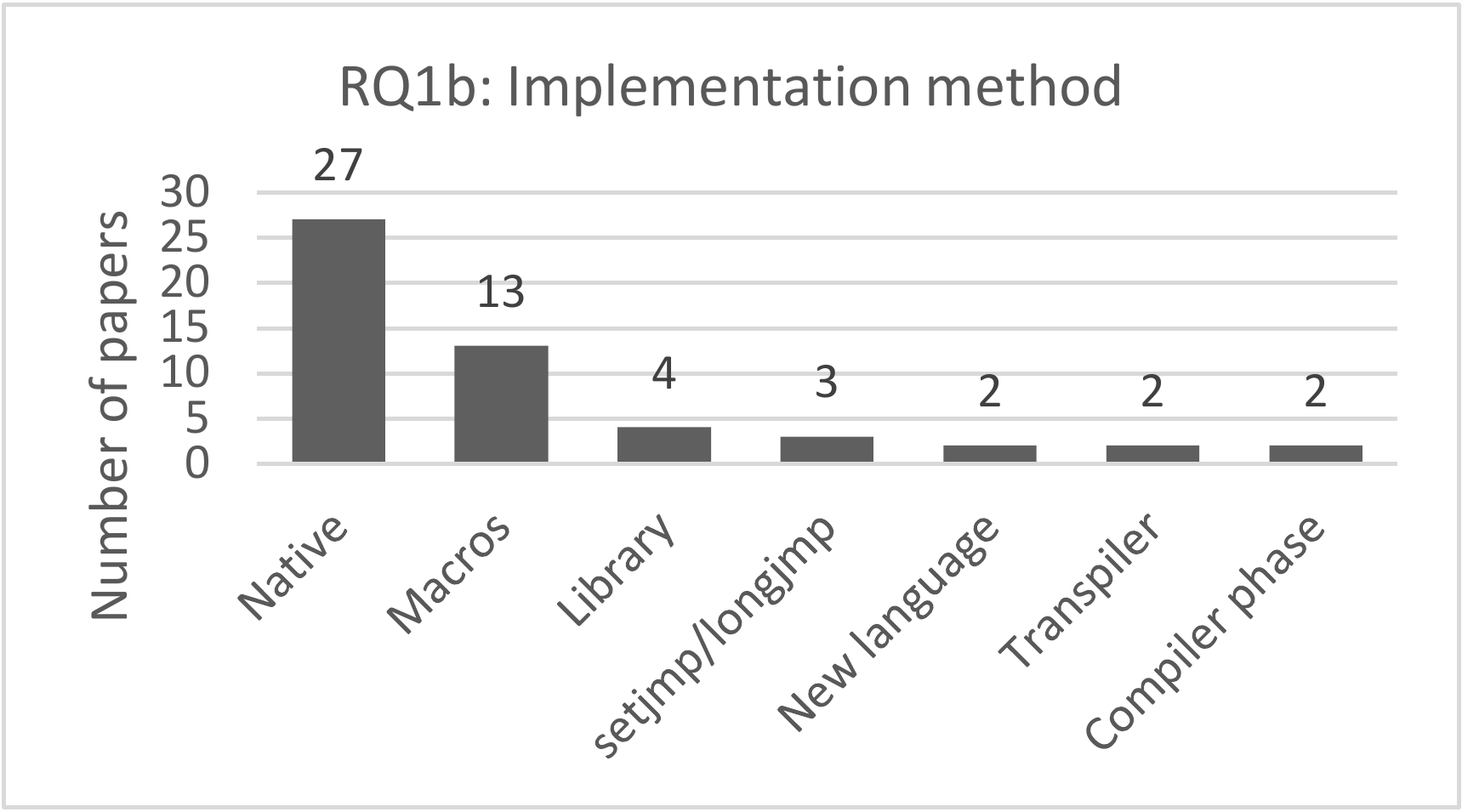}
		\caption{RQ1b - Coroutine implementation}
		\label{fig:rq1b}
	\end{subfigure}
	\caption{Language outcomes}
\end{figure}

C was the predominant programming language, as shown in Figure \ref{fig:rq1a}. 20 of the papers (57\%) used C, and a further 5 papers (14\%) used related languages (C++ and NesC). Lua was the next most common language used, with 4 papers.

\subsection{Coroutine implementation}

To implement coroutines, 27 papers (77\%) used a native method, i.e. avoiding techniques that required a new or changed tool chain. In native implementations, 13 papers employed macros (of which 7 were based on Duff's device) and 4 used libraries; in 3 papers (\cite{Yu2008, Cohen2007b, Kalebe2017}) the C setjmp/longjmp language device was used.

Several studies extended the tool chain or created a new tool. Two papers contributed new languages \cite{Jahier2016, Evers2007}, and one paper \cite{Niebert2014} provided a set of language extensions. Two papers employed a transpiler that translates from one language to another - one from Lustre to OCaml \cite{Jahier2016} and one from a synchronous extension of C to standard C \cite{Karpin2007}. One paper \cite{Fritzsche2010} used a precompiler, and one paper \cite{Jaaskelainen2008} provided a new compiler optimisation phase.

Two studies called out to another language to implement the coroutines: one \cite{Park2015}, written in the Lua language \cite{Moura2004a}, directly manipulated the hosting environment through the C API; another \cite{Khezri2008} used non-portable assembly language. The results are summarised in Figure \ref{fig:rq1b}.

\subsection{Operating system}

\begin{figure}[h]
	\includegraphics[width=0.5\textwidth]{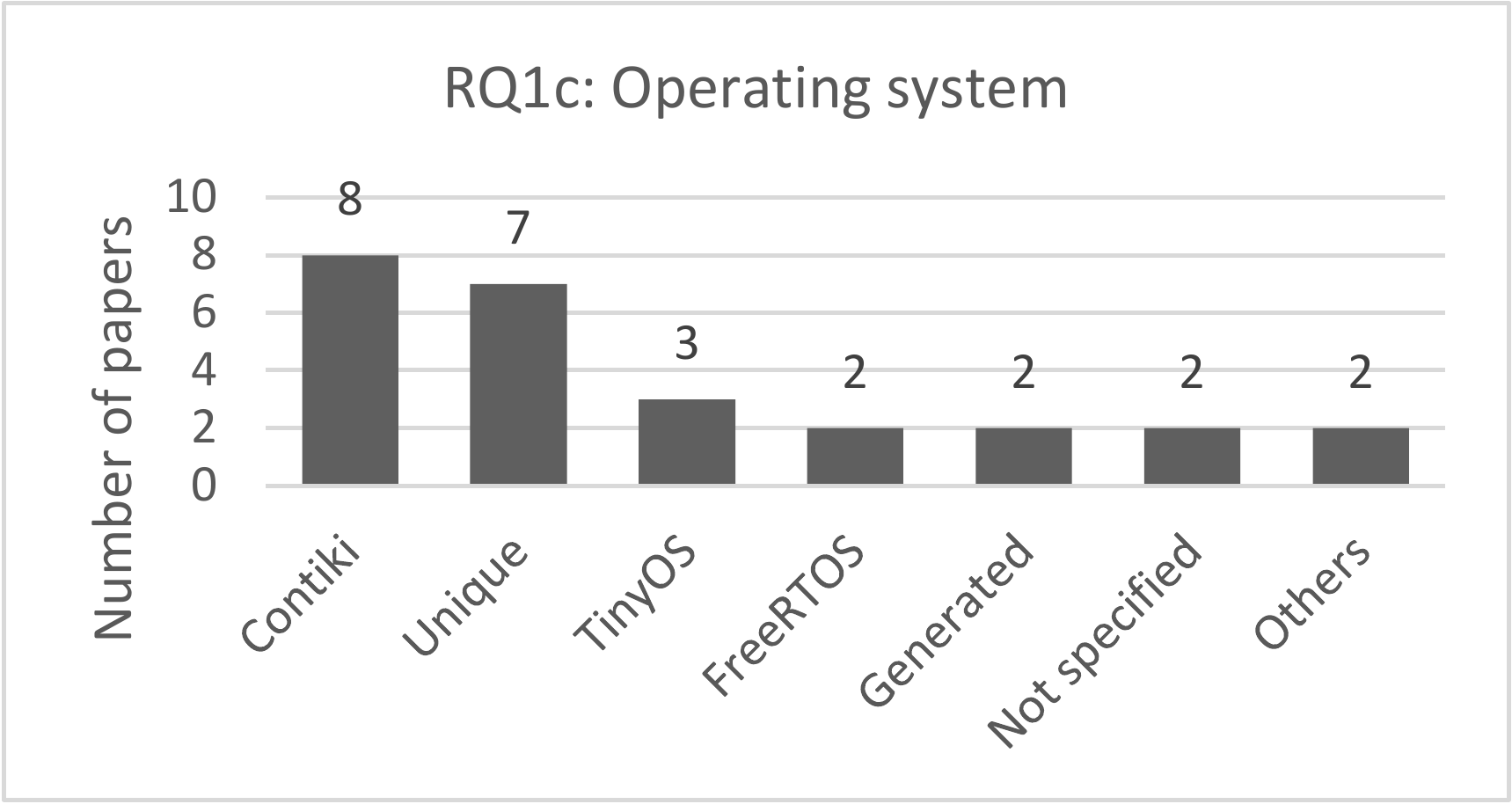}
	\caption{RQ1c - Operating systems used in selected studies using C-like languages}
	\label{fig:rq1c}
\end{figure}

Of the 26 instances studied that were written in C, C++ or NesC, 13 (50\%) used (or extended) a widely-known embedded operating system (Contiki \cite{Dunkels2004}, TinyOS \cite{Levis2005} or FreeRTOS \cite{Barry2018}) and 9 (35\%) used a unique operating system, or one that was generated for each application, as shown in Figure \ref{fig:rq1c}. There was not enough information in the papers themselves to judge how many of these 9 papers could be considered 'bare-metal'.

\subsection{Memory}

\begin{figure}[h]
	\centering
	\begin{subfigure}[h]{0.46\textwidth}
		\centering
		\includegraphics[width=\textwidth]{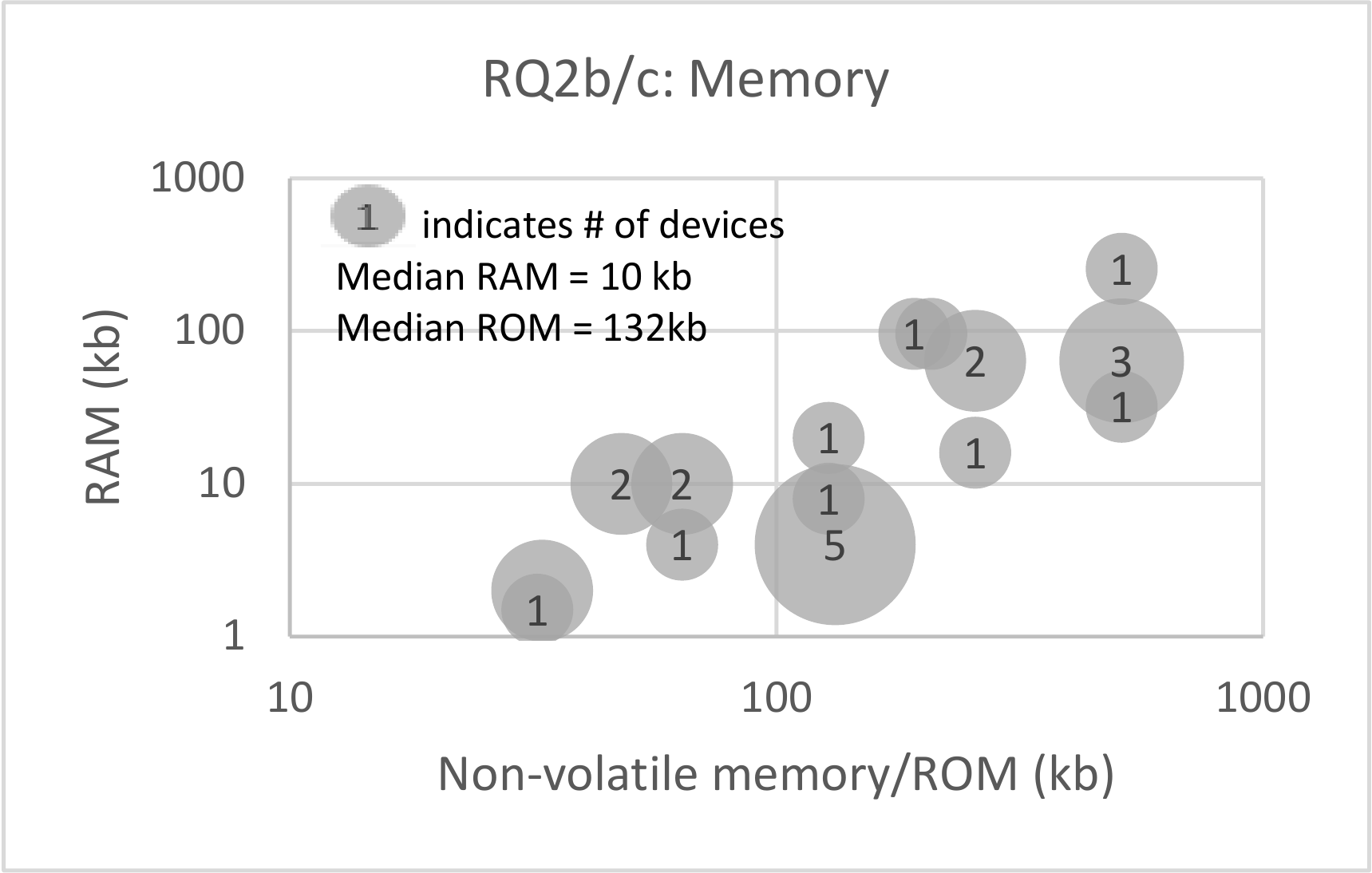}
		\caption{RQ2b/c - Memory}
		\label{fig:rq2bc}
	\end{subfigure}
	\begin{subfigure}[h]{0.46\textwidth}
		\centering
		\includegraphics[width=\textwidth]{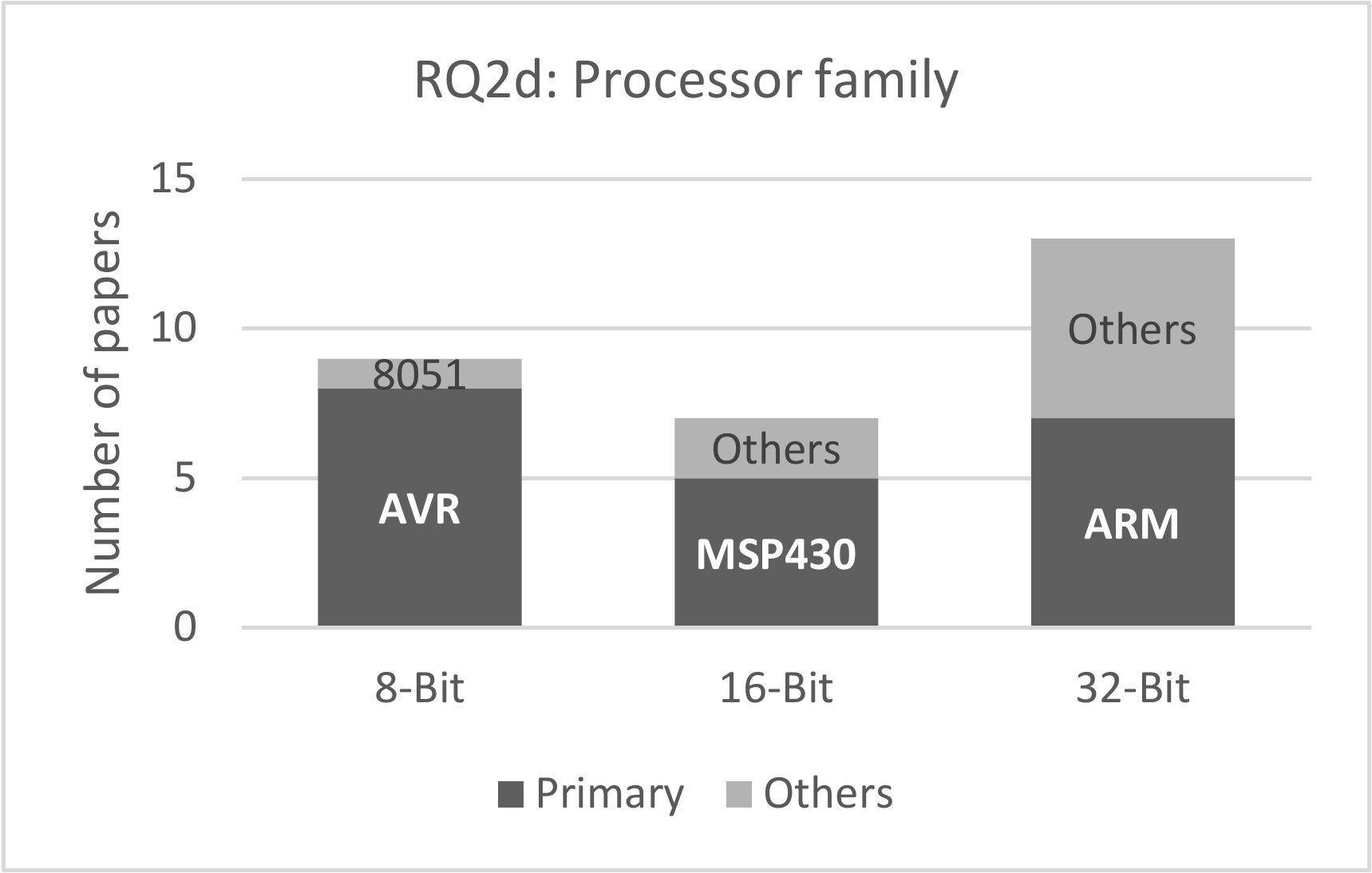}
		\caption{RQ2d - Processor families arranged by bits}
		\label{fig:rq2d}
	\end{subfigure}
	\caption{Hardware outcomes}
\end{figure}

Figure \ref{fig:rq2bc} shows the ROM and RAM sizes of the selected platforms, using logarithmic scales. As observed in RQ2a, there were many systems with low RAM sizes: the median value was 10 kb. There was a positive correlation (r=0.64) between ROM size and RAM size.

\subsection{Processors}

Only 45\% (13 out of 29) of the CPUs that were identified were 32-bit processors: 9 were 8-bit and 7 were 16-bit. The fact that 55\% (16 out of 29 studies) used 8- and 16-bit devices indicates that coroutines are applicable to very constrained platforms.

It is also notable that within the 8-bit segment, all but one were of the megaAVR family; among 16-bit processors 5 out of 7 used the TI MSP430 architecture. Within the 32-bit segment the picture was less clear-cut: just over half used the ARM architecture. These types of microcontrollers are widespread in IoT and embedded systems \cite{AspenCore2017}. These results are summarised in Figure \ref{fig:rq2d}. Full details are in the supplementary materials.

\subsection{Use cases}

\begin{figure}[h]
	\centering
	\begin{subfigure}[h]{0.46\textwidth}
		\centering
		\includegraphics[width=\textwidth]{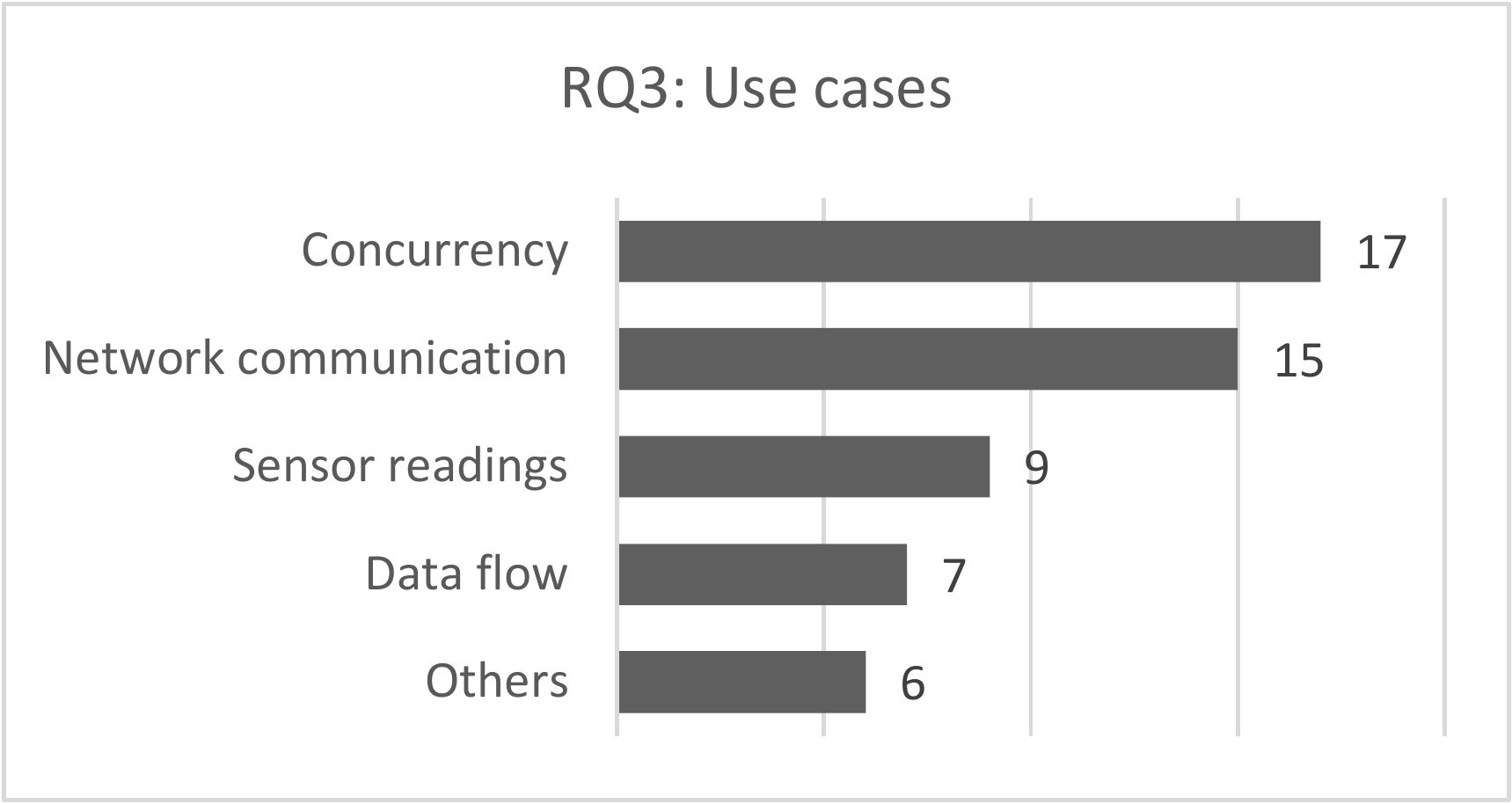}
		\caption{RQ3 - Use cases}
		\label{fig:rq3}
	\end{subfigure}
	\begin{subfigure}[h]{0.46\textwidth}
		\centering
		\includegraphics[width=\textwidth]{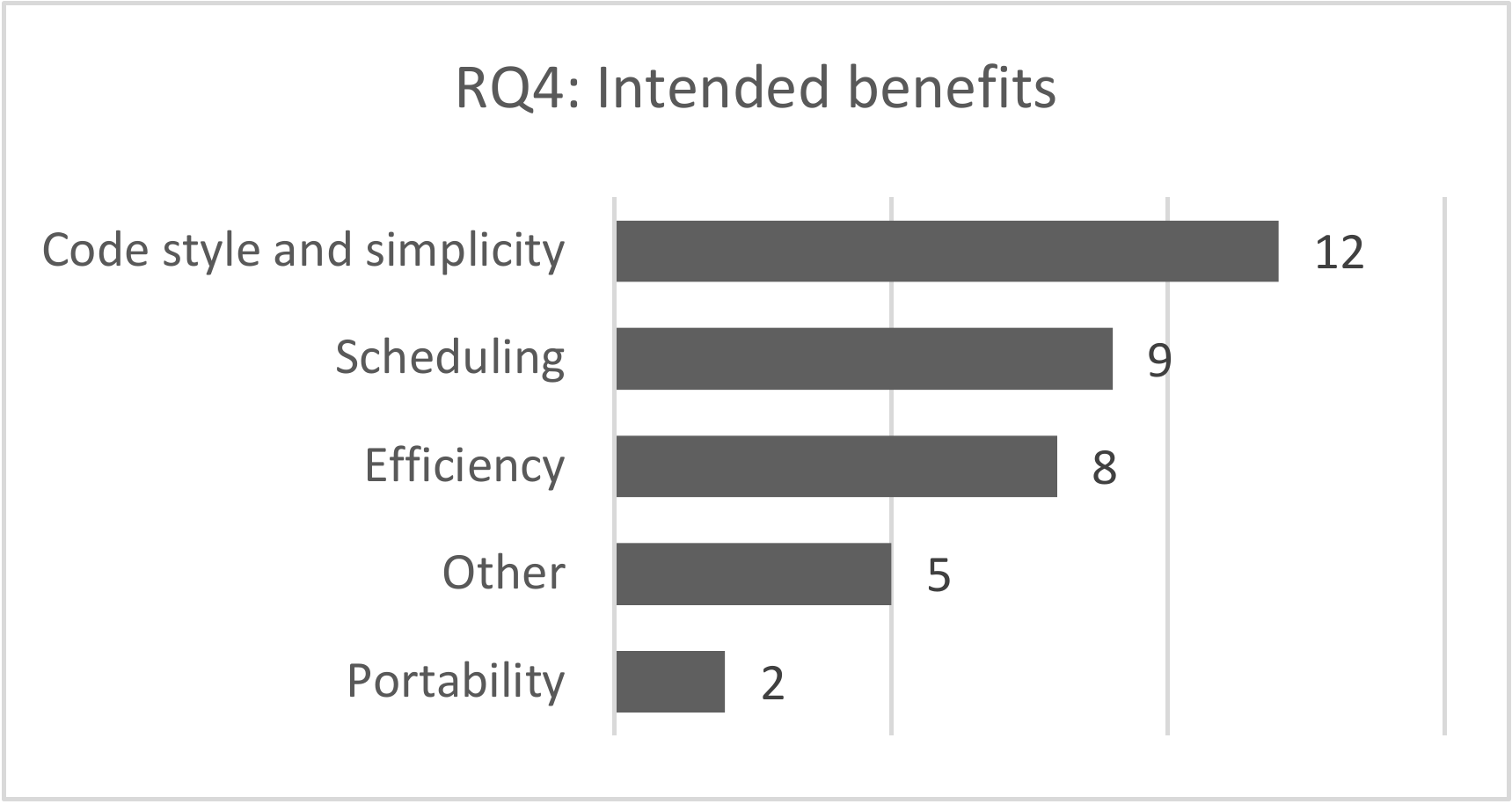}
		\caption{RQ4 - Intended benefits}
		\label{fig:rq4}
	\end{subfigure}
	\caption{Usage outcomes}
\end{figure}

The four most common use cases were concurrency (49\% of papers), network communication (43\%), sensor readings (26\%) and data flow (20\%), as illustrated in Figure \ref{fig:rq3}. It is notable that all four of these use cases are often considered to present difficulty or complexity for programmers. (See the supplementary materials for details of use cases and their classifications.)

These use cases are common across many platforms, and not just resource-constrained devices. Syntax designed for desktop systems is likely to handle these cases relatively well. Contrasting these use cases with those found in desktop development, we observe that user interfaces (a strong driver of coroutines in desktop and portable system development) are absent and that sensor readings (a rare requirement in desktop systems) are prominent.

\subsection{Intended benefits}

Of the intended benefits the most common classifications were (i) code style and simplicity (34\%), (ii) scheduling (26\%) and (iii) efficiency (23\%), as summarized in Figure \ref{fig:rq4}. (The supplementary materials contain details of the classifications of benefits.)

We have observed that split-phase programming leads to error-prone, hard-to-maintain code; it is therefore unsurprising that code style and simplicity leads the list.

However, the popularity of scheduling as a benefit of a coroutine implementation is not mirrored in mainstream desktop programming, and it may therefore not figure high in the priorities of the C++ language specification process. Coroutines provide a tool with which to build schedulers, and many embedded software applications must provide their own scheduler, either because of the special requirements of the device \cite{Inam2011, Park2015, Susilo2009} or to minimize code size by providing only the minimum requirements.

The high incidence of efficiency as an intended benefit also reflects the latency constraints of embedded systems.

\subsection{Application programming interface}

Of the 35 studies analysed, 21 discussed an implementation of coroutines: the questionnaire results for RQ5 are listed in the supplementary materials.

The API questions (RQ5b-e) could not in all cases be answered directly from inspection of the papers. In these cases, unless the answer could be found in the supplementary materials, linked source code, or was well-known to the researchers, the question was answered 'Unknown'. In some cases the source code referenced by the paper was no longer available.

\begin{figure}[h]
	\centering
	\begin{subfigure}[h]{0.46\textwidth}
		\centering
		\includegraphics[width=\textwidth]{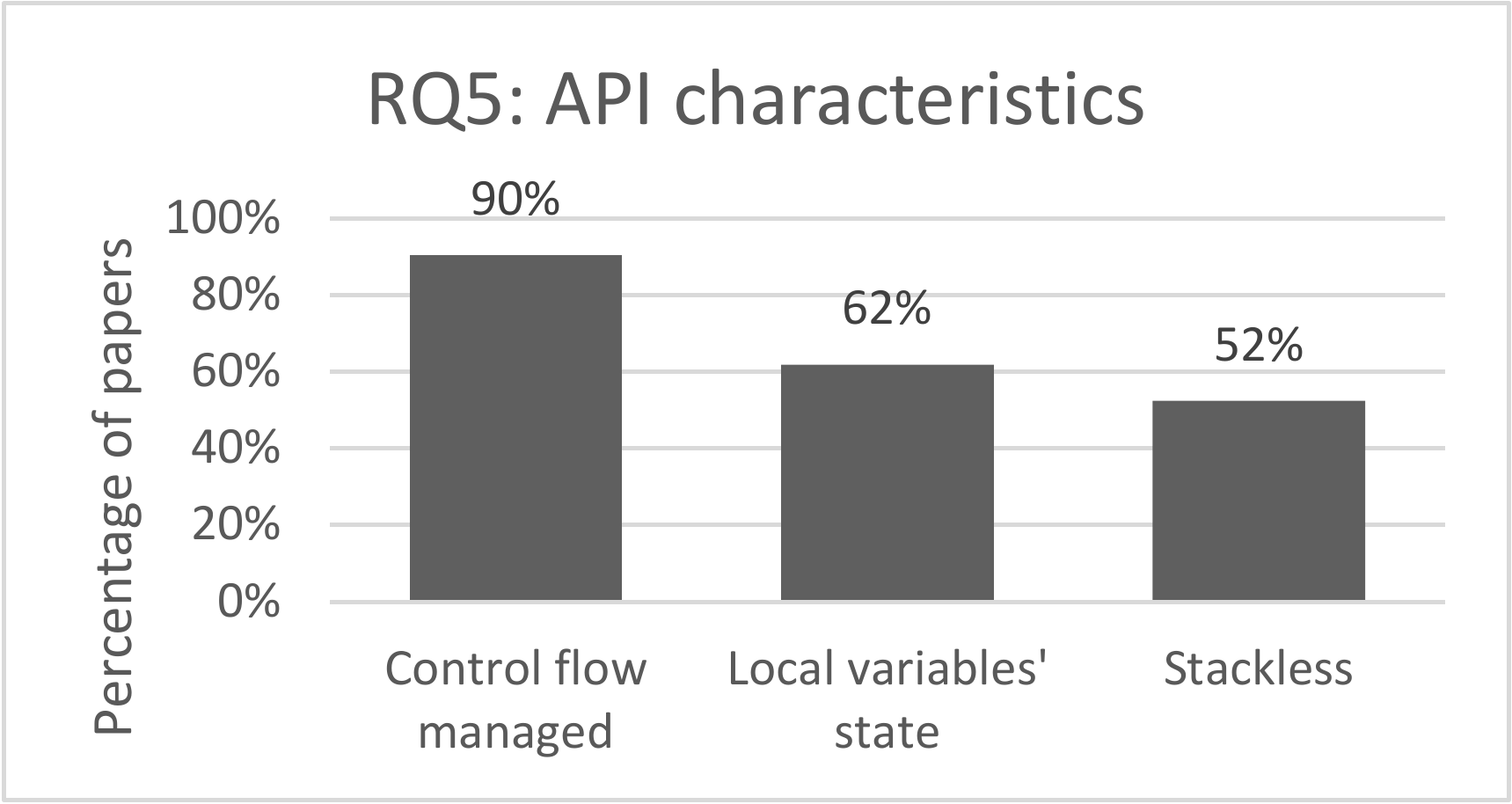}
		\caption{RQ5 - API characteristics}
		\label{fig:rq5}
	\end{subfigure}
	\begin{subfigure}[h]{0.46\textwidth}
		\centering
		\includegraphics[width=\textwidth]{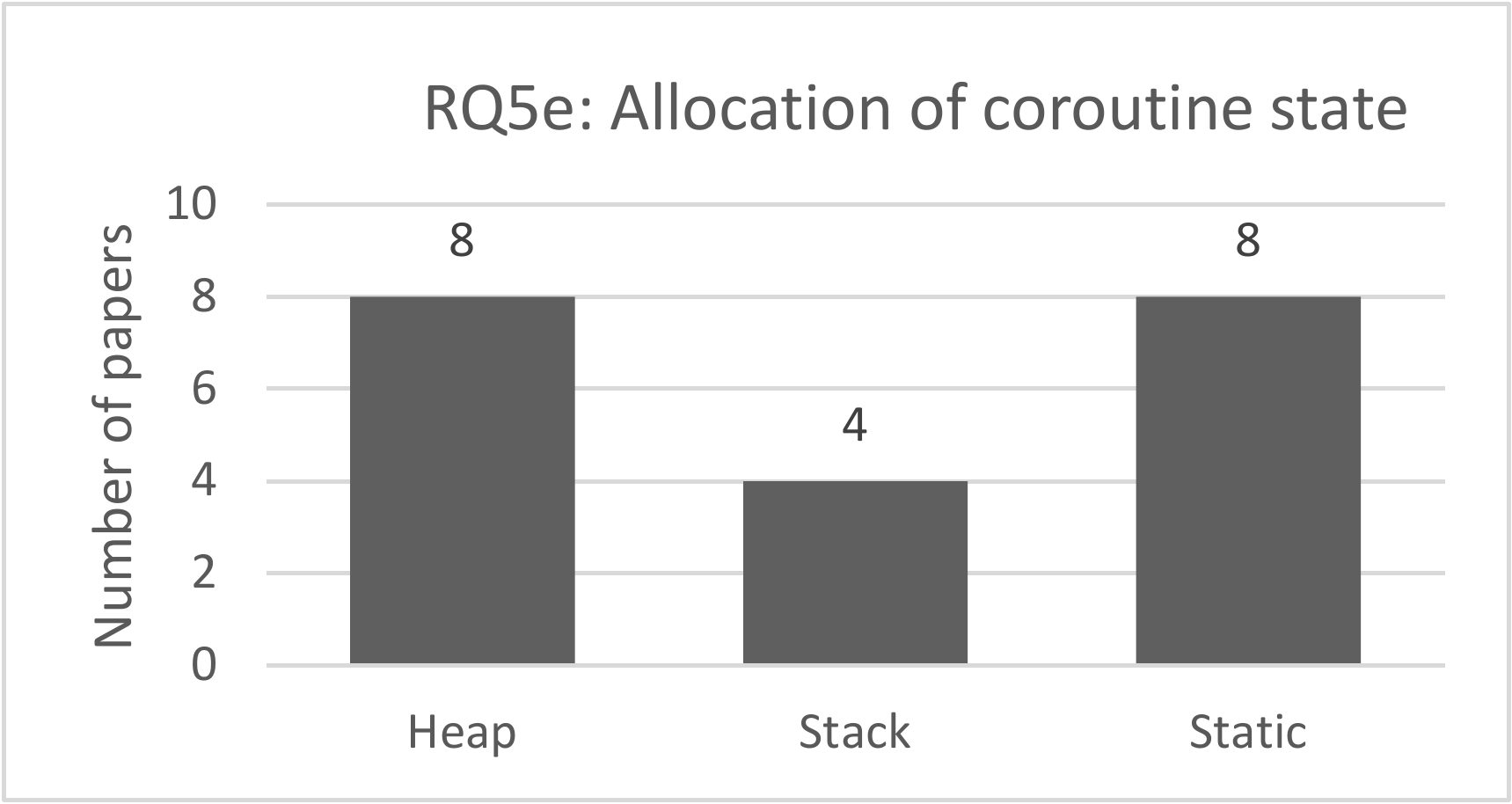}
		\caption{RQ5e - Allocation of coroutine state}
		\label{fig:rq5e}
	\end{subfigure}
	\caption{API outcomes}
\end{figure}

Figure \ref{fig:rq5} summarises the basic API characteristics for those studies which examined an implementation of coroutines. The overwhelming majority (89\%) of implementations managed control flow on behalf of the programmer; more than two-thirds managed the state of local variables. The outcome with regard to stackless and stackful implementations was more balanced: 11 stackless versus 8 stackful.

We have observed that managed, deterministic use of memory is a common requirement for embedded systems: in Figure \ref{fig:rq5e} we see that over a third of papers (8 of 21) supported the allocation of coroutine state on the heap, which is not appropriate for embedded systems, and 4 used the stack, which may not be appropriate if the state size is large or of a size unknown at compile time.

\section{Analysis and discussion}
\label{section:analysis}

\subsection{Analysis of API design}
\label{analysis-of-api-design}

\begin{figure}[h]
	\includegraphics[width=0.55\textwidth]{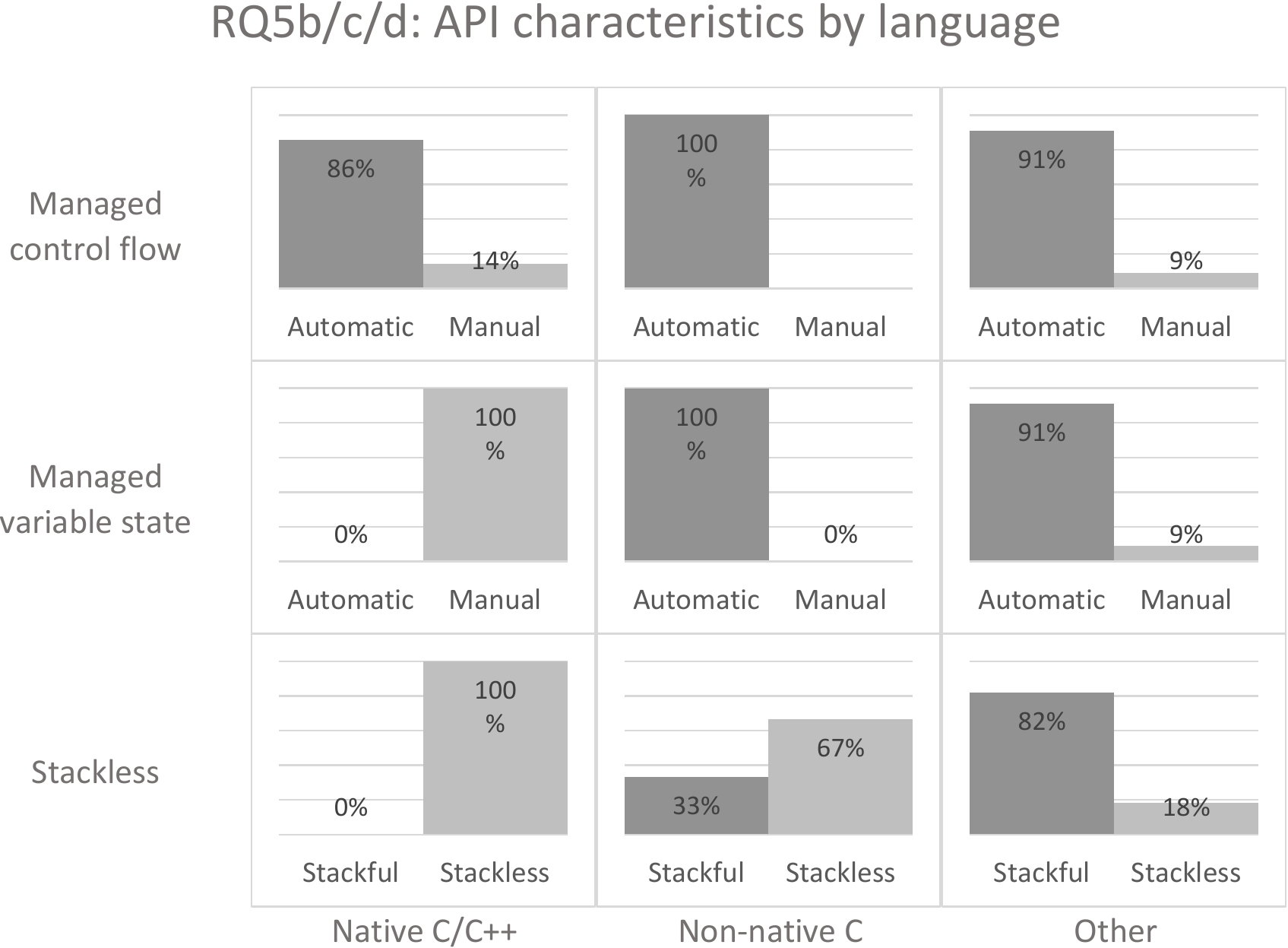}
	\caption{RQ5b/c/d - API characteristics by language}
	\label{fig:rq5bcd}
\end{figure}

%

Figure \ref{fig:rq5bcd} examines the API characteristics of the various implementations, grouped into (i) native C/C++, (ii) non-native C and (iii) languages other than C, where non-native C refers to language extensions, transpilers, or tools that otherwise extend the C compiler tool chain. The results for languages other than C and for non-native C implementations are interesting because they may reveal what the language designers and implementers considered to be desirable characteristics. (In each case the percentage shown is a fraction of the unique implementations inspected; it is not necessarily representative of the population at large.)

This paper has suggested that the management of control flow on behalf of the programmer (RQ5b) is a desirable feature of programming languages on resource-constrained platforms. The results in Figure \ref{fig:rq5bcd} appear to support this claim. All non-native C and almost all non-C implementations provide support for managing control flow. (The only exception is found in the work of \citeN{Motika2015}, a pattern whose code is primarily designed as a target for code generators.) Additionally, 86\% of the native C cases were able to provide this feature, primarily through macros.

The management of the state of a coroutine's local variables (RQ5c) has also been proposed as a desirable characteristic.  Once again, all non-native C and almost all non-C implementations provide support for this feature. None of the native C implementations were able to provide it, as a consequence of the language's limitations.

None of the native C implementations and only one of the non-native C implementations were stackful. By contrast, 82\% of the non-C implementations were stackful. It could be argued that this split indicates that, while stackfulness is a desirable feature for language designers in general, it is less desirable for C developers. Our interpretation is that, because of the perceived costs of stackfulness in terms of memory and speed, there remains strong support in the C/C++ developer community for stackless coroutines \cite{Dunkels2006}.

The allocation of coroutine state is an important feature of the design with regard to its effect on resource-constrained platforms, since it must be controlled carefully if the design is to offer predictable and safe behaviour. Of the 16 implementations where we were able to determine the allocation method, nearly a third used an object or structure to store the state. 44\% (7 instances) required that the state be allocated in static (global) memory; 1 used only the stack, and 3 offered flexibility as regards the location.

\begin{figure}
	\includegraphics[width=0.75\textwidth]{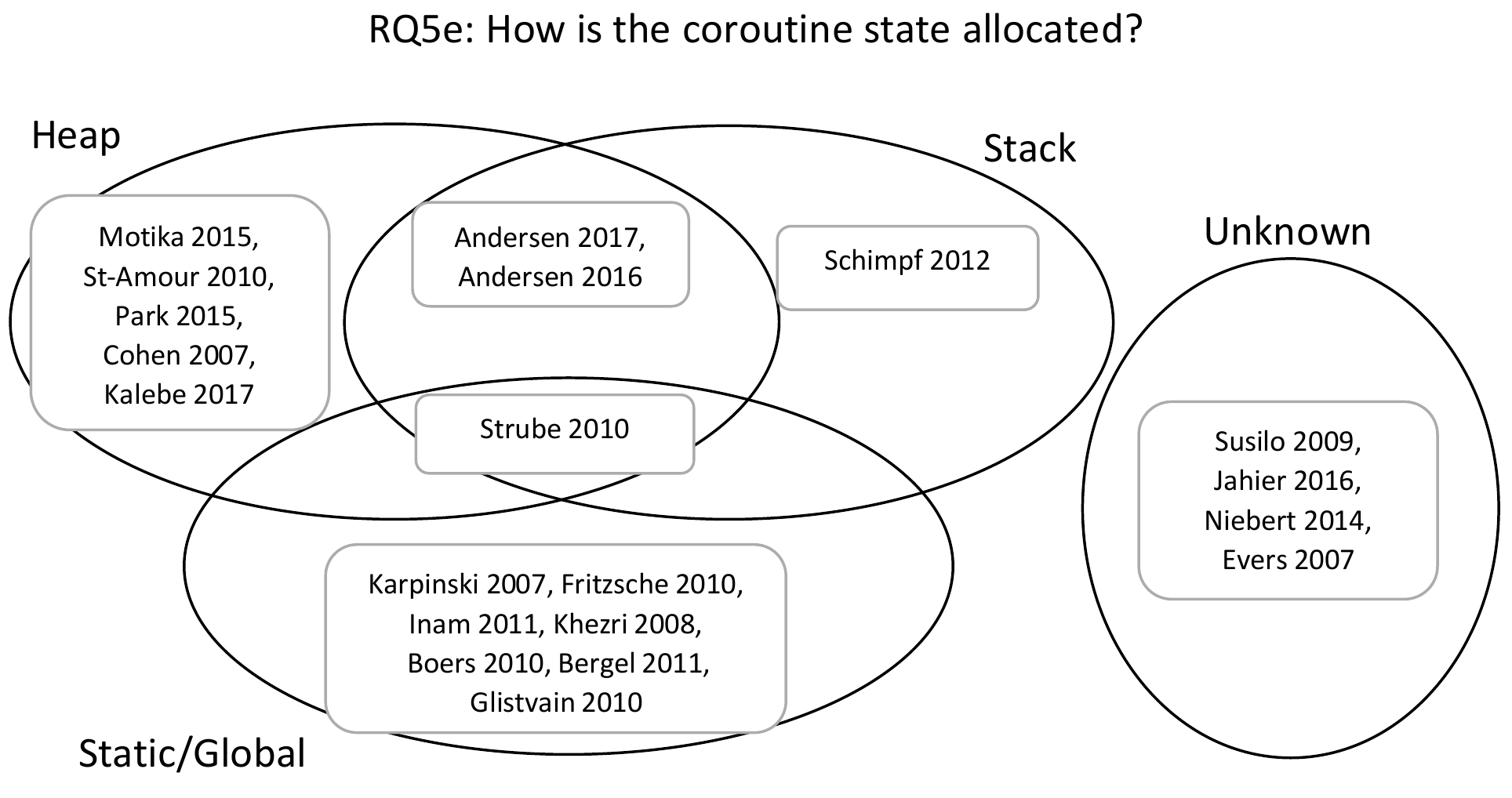}
	\caption{RQ5e - How is the coroutine state allocated?}
	\label{fig:rq5e2}
\end{figure}

5 studies (\cite{Motika2015, St-Amour2010, Park2015, Cohen2007b, Kalebe2017}) required that the state be stored on the heap (i.e. in dynamically allocated memory space).  Of these, 3 were in languages that required such a strategy (Java, Scheme and Lua) and only two (\cite{Cohen2007b, Kalebe2017}) used a C-based language (NesC or C++). In the case of \cite{Cohen2007b}, each coroutine stack of 256 bytes was allocated on the heap. However, the total number of coroutine stacks required was known in advance, and a safe allocation strategy was therefore feasible. Figure \ref{fig:rq5e2} summarises these memory strategies.

Given that mainstream C++ programming supports environments where heap memory is generally plentiful, any standard implementation of coroutines in C++ must support dynamic memory allocation for coroutine state storage. However, the special case of resource-constrained platforms, including embedded systems, requires that the developer have the option to use stack memory or global static memory, and that they have full control over which is used on each instantiation. An implementation that supports all three strategies, and allows control over which is used, is therefore desirable.%

\subsection{Research gaps}

\newenvironment{vheader}
{ \begin{sideways}\centering\begin{tabular}{c} }
{ \end{tabular}\end{sideways} }


\begin{table}[h]
	\caption{Summary of research gaps}
	\label{gaps}
	\centering
	\begin{tabular}{ m{3cm} m{1.1cm} m{1.1cm} m{1.1cm} m{1.1cm} m{1.1cm} m{1.1cm} m{1cm} }
		\hline
		  Paper
		& \begin{vheader}Language was C/C++\end{vheader}
		& \begin{vheader}Predictable memory usage\end{vheader}
		& \begin{vheader}Integrates with other \\ language features\end{vheader}
		& \begin{vheader}Maintainability \& readability\end{vheader}
		& \begin{vheader}Labour cost of implementing \\ the infrastructure\end{vheader}
		& \begin{vheader}Memory and processing cost \\ of infrastructure\end{vheader}
		& \begin{vheader}Continued use of legacy \\ code\end{vheader} \\
		\hline
Cohen et al. 2007	 & 	 & \checkmark\checkmark	 & \checkmark\checkmark	 & \checkmark\checkmark	 & \checkmark\checkmark	 & \checkmark\checkmark	 &   	\\ \hline
Evers et al. 2007	 & 	 & \checkmark	 & 	 &   	 &   	 & \checkmark\checkmark	 &   	\\ \hline
Karpinski et al. 2007	 & 	 & \checkmark\checkmark	 & \checkmark\checkmark	 & \checkmark\checkmark	 & \checkmark\checkmark	 & \checkmark\checkmark	 & \checkmark\checkmark	\\ \hline
Kumar et al. 2007	 & \checkmark\checkmark	 & \checkmark	 & 	 &   	 &   	 & \checkmark\checkmark	 & \checkmark\checkmark	\\ \hline
Khezri et al. 2008	 & 	 & 	 & \checkmark\checkmark	 &   	 &   	 & \checkmark 	 & \checkmark\checkmark	\\ \hline
Susilo et al. 2009	 & \checkmark\checkmark	 & 	 & 	 &   	 &   	 & \checkmark\checkmark	 &   	\\ \hline
Boers et al. 2010	 & \checkmark\checkmark	 & 	 & \checkmark	 &   	 &   	 & \checkmark\checkmark	 &   	\\ \hline
Fritzsche et al. 2010	 & \checkmark\checkmark	 & 	 & \checkmark\checkmark	 &   	 &   	 &   	 &   	\\ \hline
Glistvain et al. 2010	 & \checkmark\checkmark	 & \checkmark\checkmark	 & 	 &   	 & \checkmark\checkmark	 & \checkmark\checkmark	 &   	\\ \hline
St-Amour et al. 2010	 & 	 & 	 & \checkmark\checkmark	 & \checkmark\checkmark	 & \checkmark\checkmark	 & \checkmark\checkmark	 &   	\\ \hline
Strube et al. 2010	 & \checkmark\checkmark	 & \checkmark\checkmark	 & \checkmark	 &   	 &   	 &   	 &   	\\ \hline
Bergel et al. 2011	 & 	 & \checkmark\checkmark	 & 	 & \checkmark\checkmark	 &   	 & \checkmark\checkmark	 &   	\\ \hline
Inam et al. 2011	 & \checkmark\checkmark	 & 	 & 	 &   	 & \checkmark 	 & \checkmark\checkmark	 & \checkmark 	\\ \hline
Schimpf 2012	 & \checkmark\checkmark	 & 	 & \checkmark	 & \checkmark 	 & \checkmark\checkmark	 & \checkmark\checkmark	 & \checkmark 	\\ \hline
Niebert et al. 2014	 & \checkmark\checkmark	 & \checkmark\checkmark	 & 	 &   	 &   	 & \checkmark\checkmark	 &   	\\ \hline
Motika et al. 2015	 & 	 & 	 & \checkmark\checkmark	 & \checkmark 	 &   	 & \checkmark\checkmark	 &   	\\ \hline
Park et al. 2015	 & 	 & \checkmark\checkmark	 & \checkmark\checkmark	 & \checkmark\checkmark	 &   	 & \checkmark\checkmark	 & \checkmark 	\\ \hline
Andersen et al. 2016	 & 	 & \checkmark\checkmark	 & \checkmark\checkmark	 & \checkmark\checkmark	 & \checkmark\checkmark	 & \checkmark\checkmark	 & \checkmark\checkmark	\\ \hline
Jahier 2016	 & 	 & \checkmark\checkmark	 & \checkmark\checkmark	 &   	 & \checkmark\checkmark	 &   	 & \checkmark\checkmark	\\ \hline
Andersen et al. 2017	 & 	 & \checkmark\checkmark	 & \checkmark\checkmark	 & \checkmark\checkmark	 & \checkmark\checkmark	 & \checkmark\checkmark	 & \checkmark\checkmark	\\ \hline
Kalebe et al. 2017	 & \checkmark\checkmark	 & 	 & \checkmark\checkmark	 & \checkmark\checkmark	 & \checkmark\checkmark	 &   	 &   	\\ \hline
\emph{Ideal outcome}	 & \checkmark\checkmark	 & \checkmark\checkmark	 & \checkmark\checkmark	 & \checkmark\checkmark	 & \checkmark\checkmark	 & \checkmark\checkmark	 & \checkmark\checkmark	\\ \hline
\multicolumn{8}{l}{\footnotesize 
	\begin{tabular}{l}
	Key: \checkmark\checkmark - The issue is considered and resolved. \checkmark - The issue is addressed but not resolved. \\ Blank – The issue is not present. 
	\end{tabular}
}
	\end{tabular}	
\end{table}

Focussing specifically on the studies that describe an implementation, we have analysed the issues that were addressed by the research in order to identify gaps, as shown in Table \ref{gaps}. Most studies considered the memory and computational cost of the coroutine system, whereas fewer authors addressed interoperability with legacy code. The issue of predictable memory usage by coroutines is particularly important for embedded systems; although 10 of the 21 papers offered a solution, none of these solutions will apply to a C++ native solution which also handles local variable state.

We conclude that a research gap remains with regard to the study of standard C++ as an appropriate language for the development of asynchronous programs on resource-constrained devices.

\subsection{Repeatability of search results}

We found that when the IEEE Xplore database search was repeated 11 months after the original search, the new results were not, as they were expected to be, a superset of the original results. Of the original 144 papers found in October 2017, only 87 (60\%) appeared in the search results in September 2018. Further, of the 32 new results, only 16 were papers published since 2015: the other half were published before 2015. We conclude that the search methodology of the IEEE database has changed in the interim, and this raises a question over the use of this database for systematic surveys. This problem was not found for the other on-line databases used.

\subsection{Discussion}

The majority of selected papers used the C programming language; while the coroutines proposal \cite{ISO2017} is for C++, the use of C++ for these projects would not necessarily require significant programming changes.

Over a quarter of the papers used the Contiki \cite{Dunkels2004} operating system, which provides coroutine support through Protothreads \cite{Dunkels2006}, which rely on Duff's device. Given the problems, discussed in Section \ref{previous-implementations}, that are associated with using this device, this common use of Protothreads indicates a widespread need for the facilities provided by a coroutine-like solution.

More than half (55\%) of the studies used 16-bit or 8-bit processors. Support for these platforms on leading C++ compilers is currently limited; this will need to be addressed before C++ coroutines can be applied to the smallest platforms.

As expected \cite{AspenCore2017, Skerrett2017}, code style or simplicity was the leading desired benefit of the language feature implementation. The second most common benefit was a basis for a scheduler: this is not commonly a perceived benefit of coroutines on mainstream platforms, and this difference warrants further study.

The coding of three common use cases - communications, data-flow and sensor readings - present particular difficulties on constrained-resource devices, because these problems require the use of split-phase programming. Each of these problems could be addressed using programming patterns enabled by coroutines: async/await and generator. These patterns would enable a direct programming style that is likely to reduce development effort and the incidence of errors. The high incidence of these use cases in our survey indicate that they represent an important and worthwhile target for further study.

Our survey indicates that multiple studies exist that require a coroutine-based facility for concurrent programming on resource-constrained devices, establishing that a demand exists at this end of the spectrum, not merely on high performance platforms. We noted in Section \ref{analysis-of-api-design} that, where the language allowed fine-tuned management of memory allocation, dynamic allocation of memory was avoided for coroutine state and stack. We can conclude that avoiding heap memory is a requirement for the small devices that formed the bulk of the target platforms.

While 82\% of non-C implementations are stackful, only one of the C implementations is stackful. We observe that when a language is designed from the ground up to support coroutines, then a stackful implementation is common. On the other hand, such a feature is difficult in C, while preserving both backward compatibility and acceptable memory usage. We conclude that a stackless implementation is important to C programmers, and this reflects the scarcity of memory resources on the platforms under consideration.

None of the works studied utilize a coroutine implementation for C or C++ that provides managed variable state and that is designed specifically for an event-driven environment on a resource-constrained platform. We therefore conclude that this represents a significant research gap, and that further work towards such an implementation is warranted.

Although this survey found 20 papers that used C and only 2 that used C++, there is evidence that a migration from C to C++ on resource-constrained devices is occurring \cite{AspenCore2017}. Developers may also be motivated to make the switch from C to C++ to gain access to a clean implementation of coroutines to support the async/await and generator patterns and lightweight scheduling, as provided by the proposed C++20 standard \cite{ISO2017}. However, this will require language and library support appropriate for resourced constrained devices. Implementers should consider ways to avoid two of the C++ features considered dangerous by Goldthwaite \cite{Goldthwaite2006}: dynamic memory allocation and exceptions. It would be particularly useful to establish whether the proposed C++ coroutine implementations can offer deterministic memory utilisation, known at compile time: this would make it possible to avoid dynamic memory allocation.

We have seen that various specialized solutions have been applied to the problem of providing direct programming style for split-phase code on embedded systems, including Protothreads, precompilers, language extensions, post-compilation optimization phases and non-portable code libraries. It is clear that, on the one hand, coroutines offer many benefits for software development on these devices but, on the other hand, the implementation is challenging. By contrast, implementing coroutines in C++ on mainstream enterprise systems is relatively straightforward, because there are resources to spare, including memory, operating system facilities and standard libraries. While adapting coroutines for resource constrained devices may be more difficult, it offers greater benefits, because the use cases are such a good fit for embedded systems, including the low-cost, low-power scheduling, communications and sensor management that are often needed by Internet of Things applications.

\section{Conclusion and future work}
\label{section:conclusion}

\subsection{Conclusion}

This paper has analysed the current academic body of work regarding the use of asynchronous programming techniques in embedded systems. We conclude that there exists significant demand for these facilities. We argue that embedded systems must be considered as part of the debate around the standardisation of coroutines in C++. The C++ proposals provide an opportunity to improve the software engineering of embedded systems but only if the language facilities are useful in an extremely resource-constrained environment.

\subsection{Future work}

Future work could include the following:
\begin{itemize}
	\item Investigate whether the N4680 proposal can provide deterministic memory use, with full control over the detail of allocation and, if not, what changes would need to be made to the specification. Similarly, test whether the current implementations (Microsoft C++ 14.1 \cite{Microsoft2018} and LLVM 7.0.1 \cite{LLVMProject2018}) provide this determinism and control.
	\item Investigate whether the N4680 proposal and its implementations can work effectively in an event-driven environment on a resource-constrained platform, with and without a real-time operating system.
	\item Study the memory and performance costs of the current N4680 implementations on resource-constrained platforms with minimal or no operating system support.
\end{itemize}

\begin{acks}
The first author is supported by an Australian Government Research Training Program (RTP) Scholarship.
\end{acks}

\bibliographystyle{ACM-Reference-Format}
\bibliography{SMS-Coro}

\appendix

\section{Supplementary materials}

\begin{printonly}
	Supplementary materials are available in the online version of this paper. Data used in the survey can be found at https://github.com/bbelson2/supplementary-materials.
	
\end{printonly}

\begin{screenonly}
Data used in the survey can be found at https://github.com/bbelson2/supplementary-materials.
	
\begin{table}[h]
	\caption{On-line databases of digital libraries}
	
	\begin{tabular}{l l}
		\hline
		Library	& URL \\
		\hline
		ACM Digital Library & http://dl.acm.org \\
		IEEE Xplore & http://ieeexplore.ieee.org \\
		ScienceDirect & http://www.sciencedirect.com/ \\
		Scopus & http://www.scopus.com/ \\
		SpringerLink & https://link.springer.com/ \\
		Web of Science & http://www.webofknowledge.com \\
		\hline
	\end{tabular}
\end{table}
	
\begin{table}[h]
	\caption{Inclusion criteria}
	
	\begin{tabular} {l p{10cm} }
		\hline
		Identifier & Criterion \\
		\hline
		IC1 & The paper contains original research into the application of coroutines on resource-constrained platforms.\\
		IC1a & The application of coroutines must extend to the code on the platform itself, not merely to the simulator of the platform.\\
		\hline
	\end{tabular}
\end{table}

\begin{table}[h]
	\caption{Exclusion criteria}
	
	\begin{tabular} {l p{10cm} }
		\hline
		Identifier & Criterion \\
		\hline
		EC1 & The paper has no digital object identifier (DOI) or International Standard Book Number (ISBN).\\
		EC2 & The paper has no abstract.\\
		EC3 & The paper was published before 2007.\\
		EC4 & The paper is not written in English.\\
		EC5 & The complete paper was not available to the reviewers in any form equivalent to the final version.\\
		EC6 & The paper is an earlier version of another candidate paper. \\
		EC7 & The paper is not a primary study. \\
		EC8 & The paper does not fall into any of the selected publication classes. \\
		\hline
	\end{tabular}
\end{table}
\end{screenonly}

\end{document}